# Chapter 15

# Integration and (De-)installation

*P. Fessia\* and S. Weisz*

CERN, Accelerator & Technology Sector, Geneva, Switzerland

## 15    Integration and (de-)installation

### 15.1    Geographical distribution of HL-LHC upgrade interventions

The HL-LHC will require modifying the machine and infrastructure installations of the LHC in several points along the ring, in particular P1, P2, P4, P5, P6 and P7.

While the equipment in P1, P5, and P6 are meant to be installed during Long Shutdown 3 (LS3), the modifications and improvement in P4, P7 (as far as the superconducting links are concerned), and P2 shall be completed during LS2 and be operational for LHC Run 3. Below we list the activities required point by point according to the presently foreseen chronological order. P8 will have minor modifications during LS2, which are not considered here.

### 15.2    Point 4

P4 will be equipped with a new cryogenic plant dedicated to the RF systems (and other cryogenic equipment that might be installed in IR4) (Chapter 9 Section 9.5). The installation will require:

- surface: installation of the warm compressors systems in SU4;
- junction surface to the underground cavern via PX46;
- underground: installation of the cold box in TX46 (see Figure 15-1);
- connection between the cold box and QRL via UX45;
- QRL modification between −25 m and +25 m around IP4.

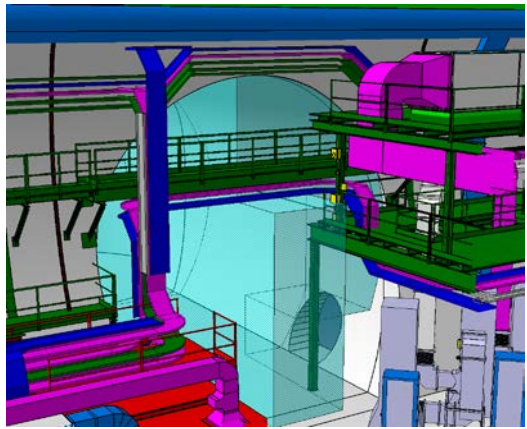

Figure 15-1: Volume reserved for the installation of the new cold box unit dedicated to RF system cooling (light blue).

---

* Corresponding author: Paolo.Fessia@cern.ch

The solutions presently adopted make the maximum use of existing volumes, requiring very limited civil engineering work, mainly devoted to the routing of the piping from SU4 to SX4 and to the vertical wall of PX46. The whole system installation shall be completed for the end of LS2. Figure 15-2 provides a schematic of the P4 buildings.

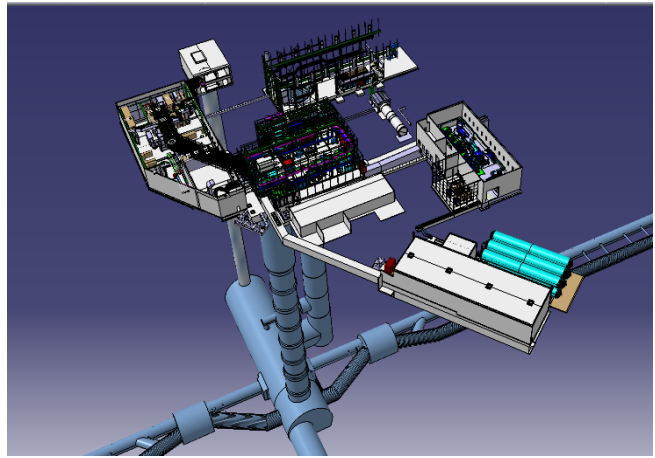

Figure 15-2: View of the LHC civil works at P4

### 15.3 Point 7

15.3.1 The horizontal superconducting links

In P7 two horizontal SC links will be installed in order to electrically feed the 600 A circuits connected to the two DFBAs (DFBAM and DFBAN). The related power converters will be installed in TZ76 and via short warm cables will be connected to the superconducting link. The two superconducting links will then run for about 220 m in TZ76 and then enter the LHC machine tunnel via UJ76. They will then be routed for about 250 m in the LHC tunnel in order to be connected to DFBAM and DFBAN (Figure 15-3). The whole installation shall take place during LS2; no long access period is available before that time. Possibilities to advance part of the interventions in the year end technical stop (YETS) shall be evaluated, and the time for radiological cooling will be used for the preparation of the interventions in the LHC tunnel. Some preparatory work has been performed already during LS, in particular the removal of the chill water pipes. This system was not of use in P7 and its dismantling has made space available for the SC link installation.

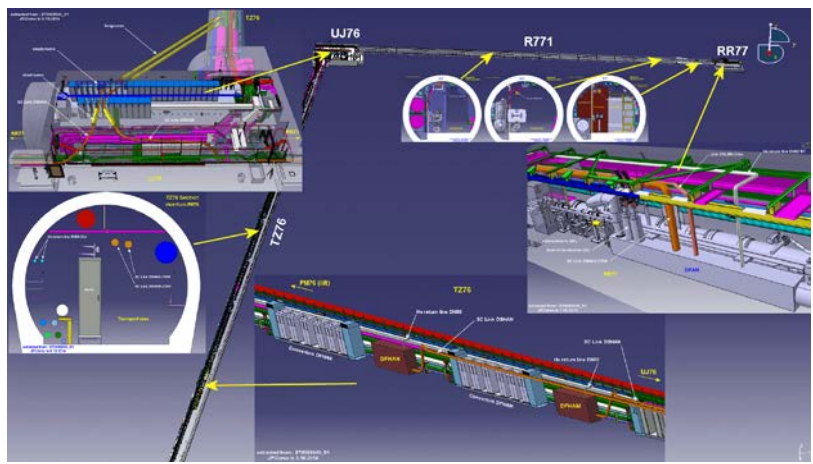

Figure 15-3: View of the foreseen installation of the superconducting link system at P7. Power converters in TZ76, routing of the link from TZ76, via UJ76 to RR77.

In order to perform the installation of the SC links in the LHC tunnel the need for de-installing part of the present equipment (as a precaution for avoiding possible damage) still needs to be fully assessed.

### 15.3.2 New collimators in the dispersion suppressor

In order to protect the superconducting magnets (excess heat deposition) from off-momentum proton leakage from the main collimator system itself, some special collimators must be installed in the dispersion suppression region, i.e. in the continuous cryostat. The evaluation of the real need for this modification will be completed based upon the first results of LHC Run 2. Unless Run 2 shows dramatic and unexpected results, the installation will eventually take place during LS3.

In order to cope with the proton losses in the dispersion suppressor area it has been decided to install two collimators on each side of the IP. In order to do so it will be necessary to:

- de-install MB.B8L7 plus MB.B10L7 and the symmetric MB.B8L7 plus MB. B10L7;
- substitute each removed dipole with a unit composed of two 11 T dipoles (Chapter 11) separated by a cryogenic bypass;
- install the collimator on the top of the cryogenic bypass (Chapter 5).

The magnet installation will also probably require a new dedicated quench protection system and a trim circuit with its own power converter. The location and the installation approach for this equipment are still under evaluation, but very probably it will be located in the nearby RR.

## 15.4 Point 2

In order to limit the heat deposition from collision debris in the superconducting magnets during the ion run, collimators in the dispersion suppressor will also be installed in P2 (Chapter 5). In this case the installation will take place only in one slot on each side of the IP replacing MB.A10L2 and MB.A10R2. This equipment is necessary for LHC Run 3 with ions, to allow Alice to fully profit from the upgrade carried out during LS2. Therefore, the installation of the unit comprising the 11 T dipoles and the cryo-bypass needs to be completed during LS2. The collimator itself can be assembled at the same time or later in a short technical stop or in a YETS. All the other issues listed in Section 15.3.2 are applicable, but the proximity services will probably be installed in the UAs.

## 15.5 Point 6

In P6 the two Q5 quadrupole magnets will be modified in order to fulfil the needs of the new HL-LHC ATS optics. Two options are presently under evaluation leading either to the exchange of the present Q5 with a new Q5 or the installation of Q5-bis near the currently operated Q5. Both options will require the local reshuffling of the vacuum layout and possible modifications to QRL.

The intervention in P6 is required to be completed in LS3, but options to anticipate it in LS2, both to limit the amount of work during LS3 and to reduce personnel exposure to radiation due to the proximity to the beam dump line, are under study.

## 15.6 Point 1 and Point 5

The largest part of the new equipment required by the HL-LHC performance objectives will be installed in P1 and P5. The items to be installed and actions to be carried out are listed below, and are applicable to both points if not otherwise specified. The list is organized by geographical areas.

### 15.6.1 LHC machine tunnel

De-installation.

- All the machine equipment from the interface with the experimental cavern (TAS included) until the DFBA (included) requires removal.
- The present QRL will be removed from the same area. A new return module will be installed in the ex-DFBA area in order to allow separation of the coolant flows coming from the LHC QRL and one of the new HL-LHC QRLs that will feed part of the machine from Q1 to Q6. This return module should also provide the possibility, if required, to connect the LHC QRL with the newly installed one, ensuring an increased level of redundancy in the system.
- Services linked to the above de-installed equipment will be removed and new ones shall be installed:

Preparation for re-installation.

- Minor works could be necessary in order to prepare the tunnel floor and wall to receive the installation of the new equipment (for larger and dedicated civil engineering works as described below).

Installation of the new equipment, probably in the following sequence:

- TAXS;
- services;
- QRL with related valve and service modules;
- horizontal superconducting links from the DFM to the magnets to be fed;
- Magnets and crab cavity support system;
- magnets and crab cavity;
- distribution feedboxes (DFX) for the Q1 to D1 magnet system and distribution feedboxes (DFM) for the D2 to Q6 magnet system.

The sequence of installation of the vertical superconducting links to be connected to the DFX and DFM still need to be assessed according to the options retained for their routing.

In addition to the interventions described above, it may be necessary to strengthen the collimation system with new collimators in the dispersion suppressor as described for P7 in Section 15.3.2.

### 15.6.2 Existing LHC tunnel service areas

The RRs on both sides of P1 and P5 will need to be re-organized, and in particular the following will be necessary.

- To de-install the power converter and other related systems (e.g. the quench detection system) linked to the powering of the removed LHC matching section.
- To re-organize the remaining equipment in order to have the most efficient space occupation, increase if necessary the radiation shielding, and place the most radiation-sensitive equipment in the most protected areas. Possible equipment replacement with new radiation-hard elements can be envisaged.

At present no civil work interventions are foreseen in the RRs areas.

### 15.6.3 New HL-LHC tunnel service areas

The installation of the new cryogenic plant in P1 and P5 will have two objectives:

- to provide independent and redundant cooling capacity to feed the final focus and matching sections left and right of each of the two high luminosity insertions for the LHC;
- to provide redundancy to the cryogenic plant installed to cool the experimental systems.

The cold boxes shall be installed in underground areas. At present the required volume does not exist. Therefore conceptual studies have started in order to identify the best options for building new underground caverns to install this equipment and the related service and control system. Figure 15-4, Figure 15-5, and Figure 15-6 correspond to solutions with power converters in the underground areas while Figure 15-7 corresponds to a solution with magnet power converters on the surface (baseline).

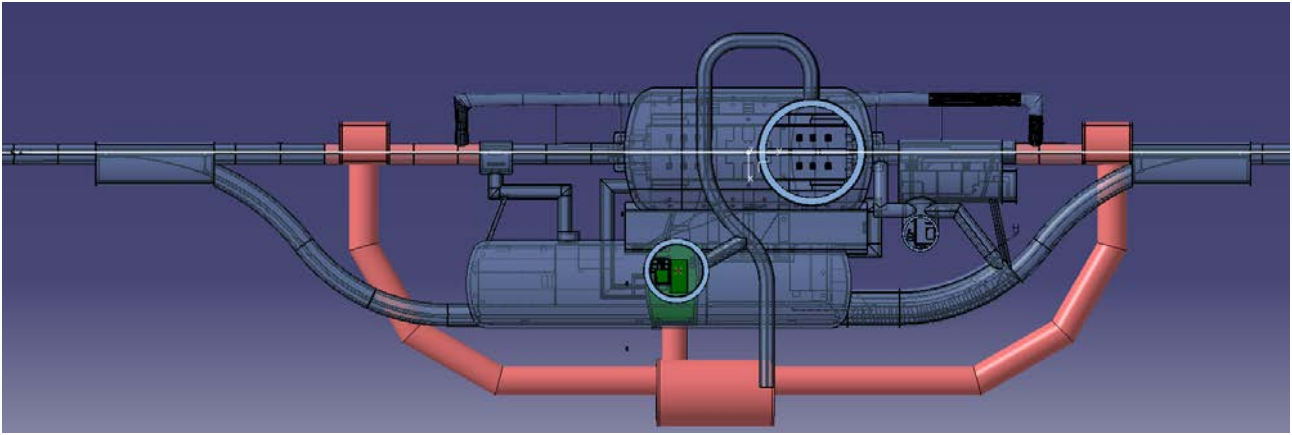

Figure 15-4: Possible option for underground work at P5 including links to the LHC tunnel. This solution is without an independent shaft to access to the surface. New structures are in light red, and zones of the LHC tunnel to be impacted by construction the new structures are in green.

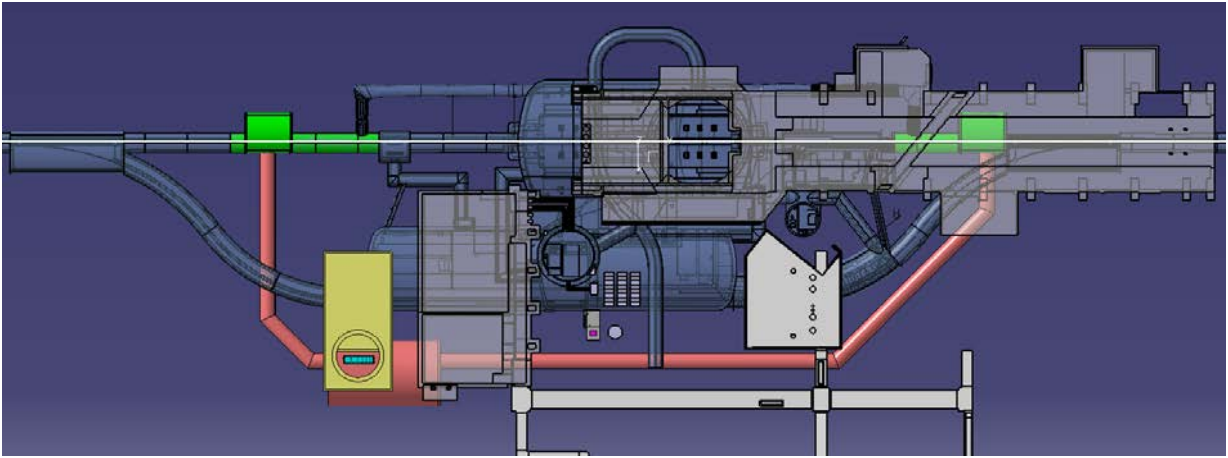

Figure 15-5: Possible option for underground work at P5 including links to the LHC tunnel. This solution has an independent shaft for access to the surface. New structures are in light red and yellow, and zones of the LHC tunnel to be impacted by construction of the new structures are in green.

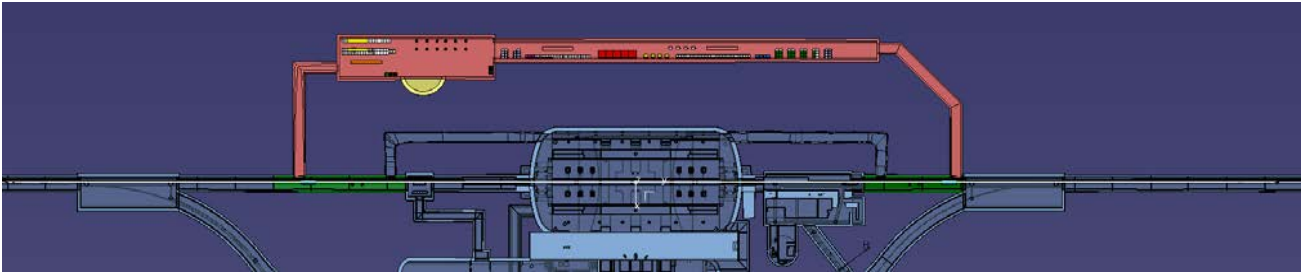

Figure 15-6: Possible option for underground work at P5 including links to the LHC tunnel. This solution has an independent shaft for access to the surface, and a larger connection tunnel for possible power converter underground installation. New structures are in light red and yellow, and zones of the LHC tunnel to be impacted by construction of the new structures are in green.

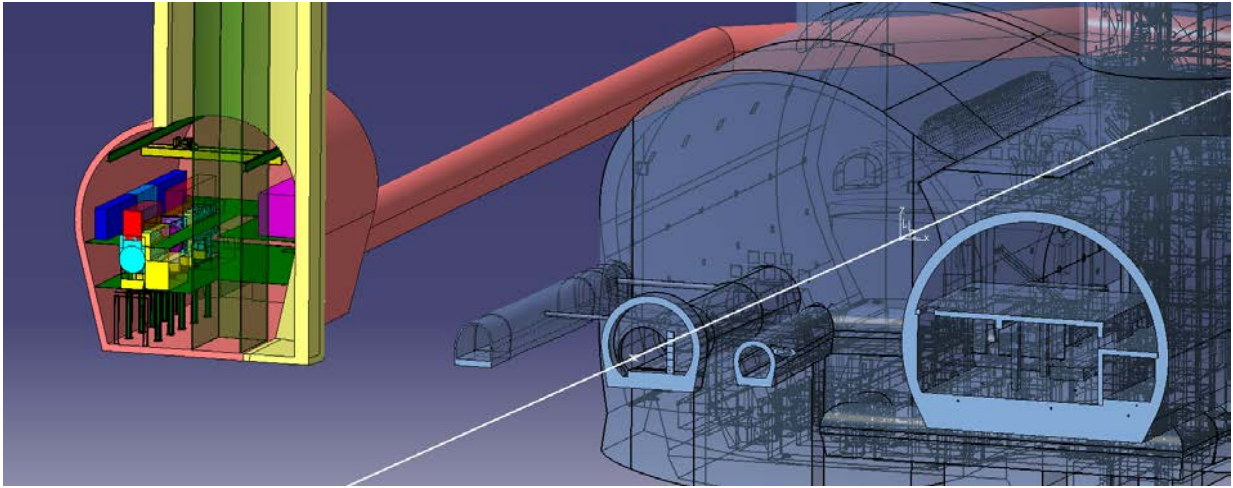

Figure 15-7: Possible option for underground work at P5 including links to the LHC tunnel. This solution has an independent shaft for access to the surface. A cross-section through the cryogenic equipment is shown. New structures are in light red and yellow.

### 15.6.4 New connection from the LHC tunnel and HL-LHC service areas to the surface

The following connections between the surface and the underground installation shall be made available.

- LHC tunnel and crab cavity area to the surface. The crab cavities need to be connected to the dedicated RF power system and their control system. The present preferred choice is to install these services in dedicated surface buildings (see below). The underground to surface connection will need to house eight RF coaxial cables (each about 300 mm in diameter), plus the required control cables. Two options are being studied.
    o The construction of eight ducts over each crab module, linking directly the tunnel vault with the surface crab service building.
    o The construction of a shaft where all the previously listed cables can be hosted together. In this case a local enlargement of the LHC tunnel will be necessary. Alternatively there will be a need for the construction of a separate small cavern where the shaft will end. The extra underground work is necessary to take into account that the minimum shaft dimension is larger than the LHC tunnel diameter.
- New HL-LHC service area to the surface. These connections are necessary to link the surface part of the cryogenic plant with the cold box installed in the new underground HL-LHC service areas.
- Vertical routing of the superconducting links. In each point at least four superconducting links will need to be routed from the surface to the underground areas. Three options are possible and they are listed from most to least convenient:
    o installation of the superconducting links in the same vertical duct linking the cryogenic plant to the surface;
    o installation of the superconducting links in the same vertical duct linking the crab cavity to the surface;
    o installation of the superconducting links in new dedicated vertical ducts.

### 15.6.5 New surface installation

The following installations shall find space on the surface for P1 and P5 installations.

- Crab cavity RF power and services hosted in two ad hoc surface buildings (yellow boxes in Figure 15–8). They shall be positioned on the surface, directly above the tunnel position where the crab cavities will be installed. There will be two surface buildings for each point, one on the lefthand part of the machine and one on the righthand part. The surface extremities of the ducts/shaft for the crab cavity coaxial leads or shaft shall be housed inside this building. The part of the building with the controls system shall be electromagnetic shielded.
- Cryogenic installation. The warm compressors and the other parts of the cryogenic plant shall be installed on the surface.
- Power converters, upper extremities of the superconducting links, protection systems, and energy extraction system related to the circuits fed via the superconducting link. This area may possibly be located near the surface part of the cryogenic plant and in any event on top of the surface routing of the vertical superconducting link.

Currently it looks very difficult to find the required space in existing buildings; therefore the construction of new structures to host the systems listed above is probably necessary (Figure 15-9).

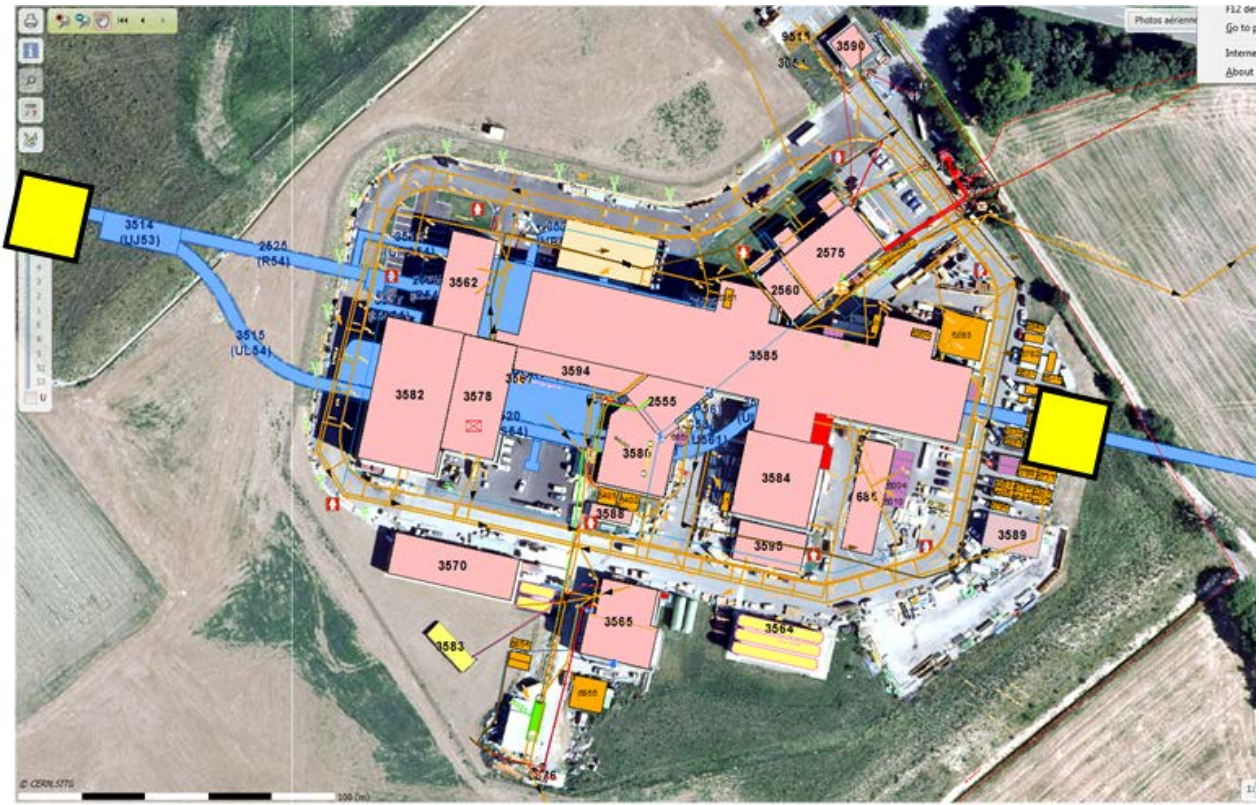

Figure 15-8: Surface position of the two new crab cavity service buildings (in yellow) at P5

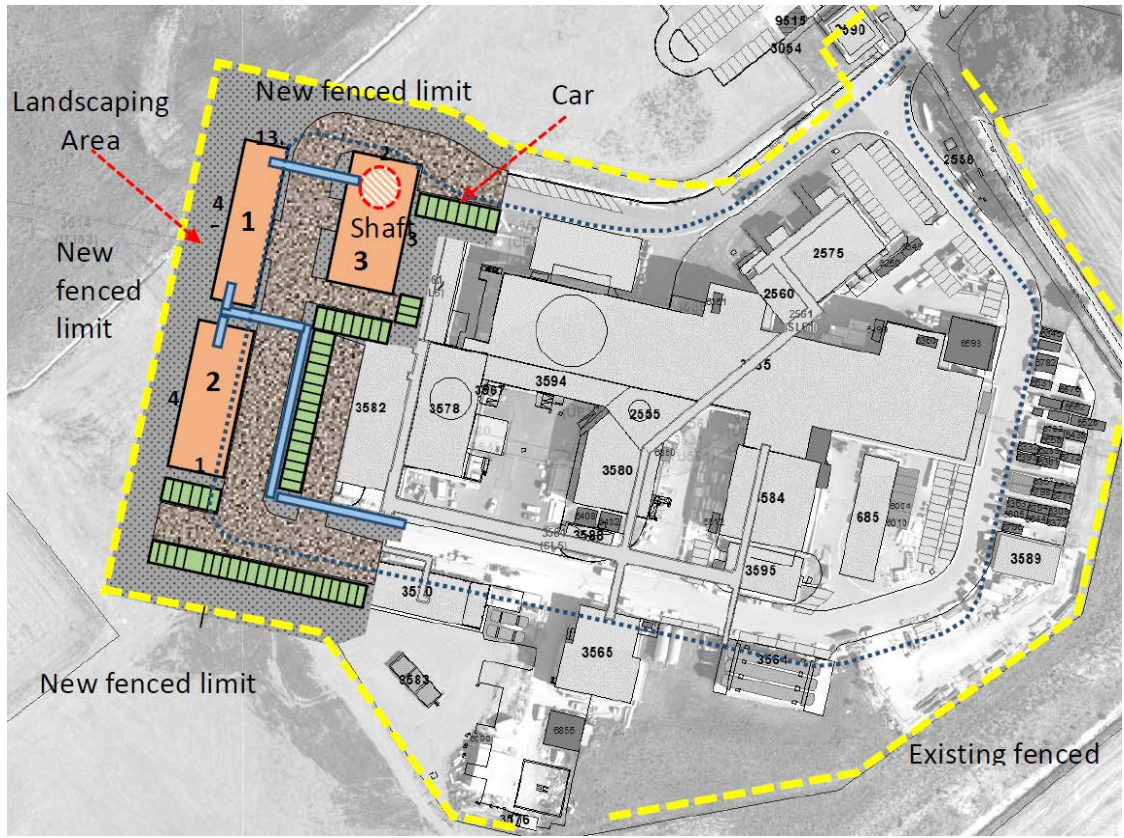

Figure 15-9: Preliminary study of the LHC P5 building footprint. Building 1: power converters and SC link terminals; Building 2: cryogenic surface equipment; Building 3: access to shaft. Possible technical galleries are in light blue.

In addition to the above mentioned options, a solution, where the power convertors and the crab cavity ancillaries would also be installed underground, is under study. Such an option would lead to an underground structure of much larger volume, but the increase in cost could be partially balanced by a reduction in the number of cores from the surface to the tunnel, a reduction in the numbers of surface structures to be built, and the simplification of the surface worksite.

### 15.6.6  Activity sequence considerations

While the sequence of intervention on the underground LHC installed equipment (machine and RR) is quite clear and linked to the end of the LHC Run 3, the sequence for the other installation and civil work activities is still under evaluation. The worksite organization will depend upon the options chosen. This is why, in the evaluation of the selection of the final solutions for the underground caverns, the connection tunnel to the surface, and the surface building, three main parameters shall be taken into account. These are cost, the possibility of performing the work in advance with respect to the LS3 start date, and the ease of installation and connection, since some options could lead to very complex and cross-connected installation worksites, making LS3 unacceptably long.